\documentclass[final,5p,times,twocolumn]{elsarticle} 

\usepackage{lineno,hyperref,graphicx,caption,subcaption}
\modulolinenumbers[200]
\journal{Nuclear Instruments and Methods in Physics Research A}

\begin{document}

\begin{frontmatter}

\title{Total Ionizing Dose Effects on CMOS Image Sensor for the ULTRASAT Space Mission}


\author[mymainaddress,otheraddress]{Vlad D. Berlea}

\author[mymainaddress,otheraddress]{Steven Worm}
\author[mymainaddress]{Nirmal Kaipachery}
\author[mymainaddress]{Shrinivasrao R. Kulkarni}
\author[mymainaddress]{Shashank Kumar}
\author[mymainaddress]{Merlin F. Barschke}
\author[mymainaddress,otheraddress]{David Berge}

\author[mysecondaryaddress]{Adi Birman}
\author[mysecondaryaddress]{Shay Alfassi}
\author[mysecondaryaddress]{Amos Fenigstein}

\address[mymainaddress]{Deutsches Elektronen-Synchrotron DESY, Platanenallee 6, 15738 Zeuthen, Germany}
\address[otheraddress] {Institut f{\"u}r Physik, Humboldt-Universit{\"a}t zu Berlin, Newtonstrasse 15, 12489 Berlin, Germany}
\address[mysecondaryaddress]{Tower Semiconductor, 20 Shaul Amor Avenue, Migdal Haemek, 2310502, Israel}

\begin{abstract}
ULTRASAT (ULtraviolet TRansient Astronomy SATellite) is a wide-angle space telescope that will perform deep time-resolved surveys  in the near-ultraviolet spectrum. ULTRASAT is a space mission led by the Weizmann Institute of Science and the Israel Space Agency and is planned for launch in 2025. The camera implements backside-illuminated, stitched pixel sensors. The pixel has a dual-conversion-gain 4T architecture, with a pitch of $9.5$ $\mu m$ and is produced in a $180$ $nm$ process by Tower Semiconductor.

Before the final sensor was available for testing, test sensors provided by Tower were used to gain first insights into the pixel's radiation tolerance. One of the main contributions to sensor degradation due to radiation for the ULTRASAT mission is Total Ionizing Dose (TID). TID measurements on the test sensors have been performed with a Co-60 gamma source at Helmholz Zentrum Berlin and CC-60 facility at CERN and preliminary results are presented.
\end{abstract}

\begin{keyword}
\texttt 4T CMOS Imaging Sensor (CIS), Tower $180$ $nm$, Backside Illumination (BSI), RTS, TID
\end{keyword}

\end{frontmatter}

\linenumbers

\section{Introduction}

\begin{figure}[h]
	\includegraphics[width=.55\columnwidth]
	{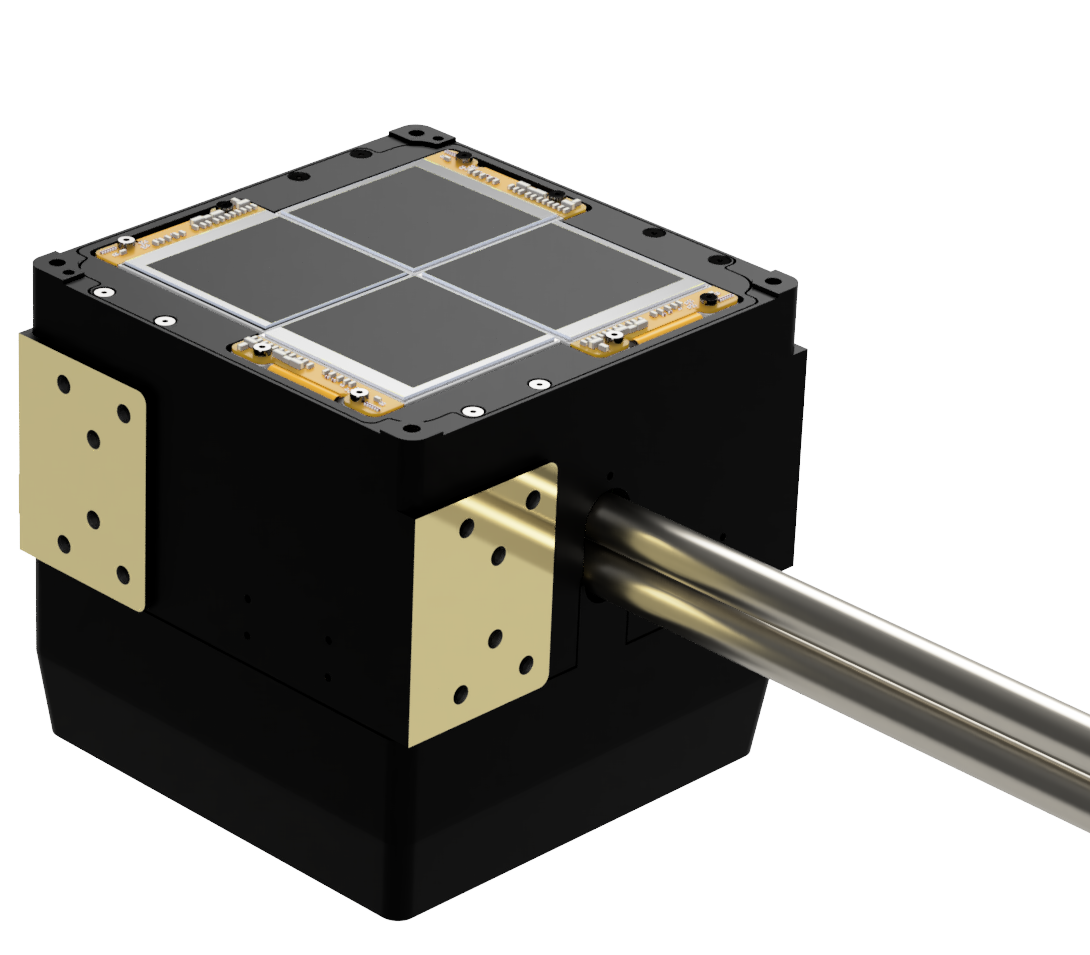}\hfill
	\includegraphics[width=.45\columnwidth]
	{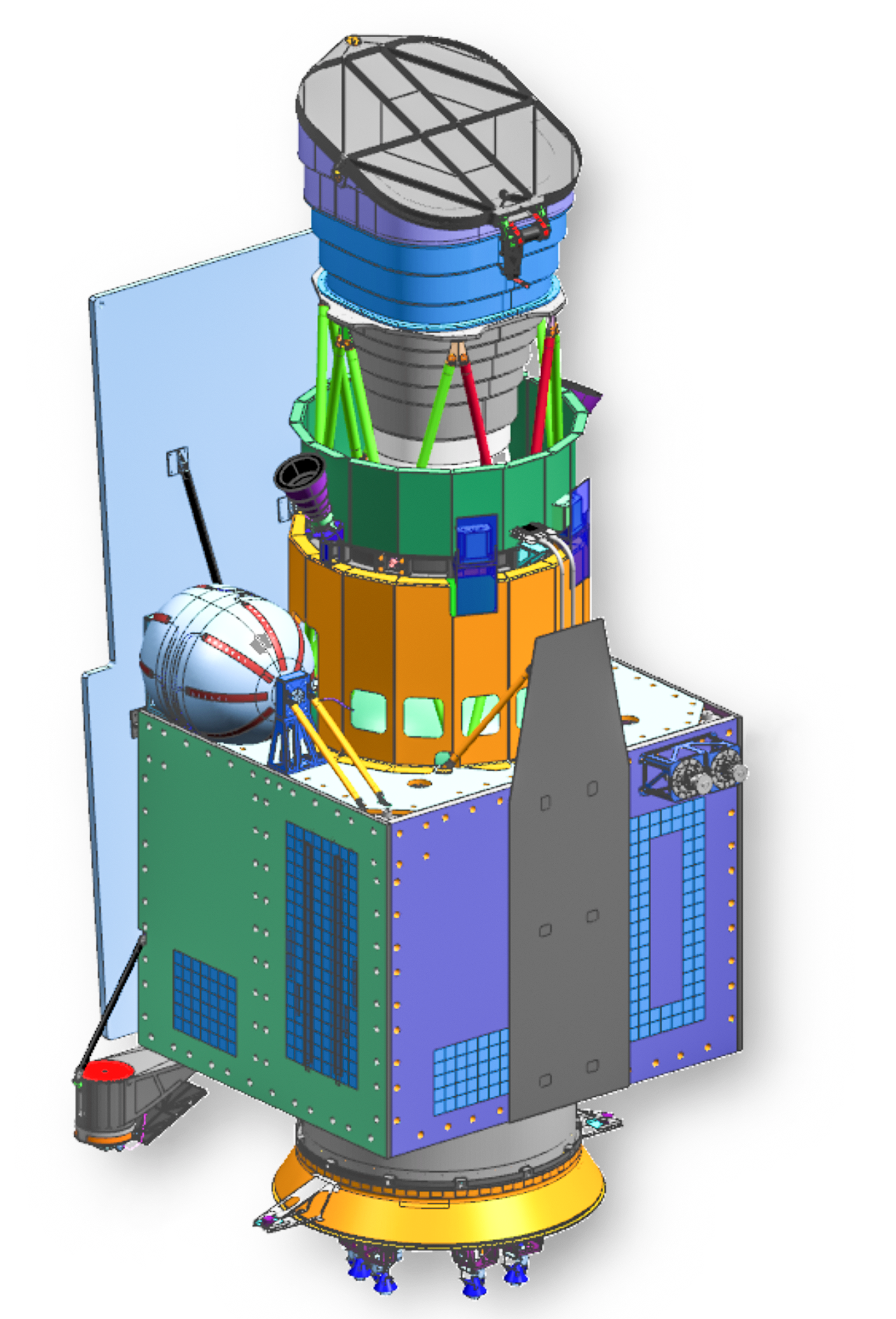}
	
	\caption{Computer rendering of the ULTRASAT camera design (left) and the ULTRASAT satellite (right). }
	\label{fig:renders}
\end{figure}

ULTRASAT is a wide-angle space telescope that will perform a deep time-resolved all-sky survey in the near-ultraviolet (NUV). The science objectives are the detection of counterparts to gravitational wave sources and supernovae. The UV camera, seen in Figure \ref{fig:renders} is composed of the detector assembly and the remote electronics unit and is being developed by DESY.

The UV camera is comprised of 4 independent CMOS sensors with a total active area of $81$ $cm^2$ and $90$ Megapixel. Each individual sensor is a CMOS imaging sensor built in the $180$ $nm$ Tower architecture. The relatively large sensor, with an active area of $4.5$ x $4.5$ $cm^2$ is achieved, through stitching multiple blocks together. A backside illumination process is employed as this is expected to yield higher quantum efficiency.

During the 3 years of operation in the geostationary orbit, the CMOS camera will be exposed to significant charged particle fluxes. These will induce Total Ionizing Dose (TID) in the sensor. A dose $< 20$ krad (including radiation design margin) is expected to be incurred by the sensor during the design lifetime of the mission

\section{Test sensors}

Before the flight design of the sensor was produced, tested and packaged, preliminary TID measurements have been performed on test sensors \cite{10.1117/12.2593897}. These test sensors are built in the same Tower $180$ $nm$ architecture with a similar collection photo-diode. Several of these test sensors have been irradiated with a Co-60 source at both HZB and CC-60 CERN. All samples have been biased (highest applicable values) and kept at room temperature during irradiation. Measurements have been performed at irregular intervals after irradiation (up to 2 months) due to significant data acquisition times. 

\section{TID impact on the dark current}

Irradiating a CMOS sensor, leads to charge traps in the SIO$_{2}$ layer and at the interface between the oxide and the epitaxial layer. These, in turn, are expected to lead to increased dark current in the photo-diode. As long as the Transfer Gate (TG) works in accumulation mode, the space charge region of the photo-diode remains insulated from the interface states near the oxide. With the accumulation of interface states, the electrical properties of the photo-diode can change and the field lines in the space charge region can be extended to the oxide. This effect is expected to be enhanced by using a non-negative bias value for the Low TG line \cite{2017.2779979}, which is the case for the ULTRASAT sensor.

\begin{figure}[h]
	\centering
	\includegraphics[width=0.81\columnwidth]
	{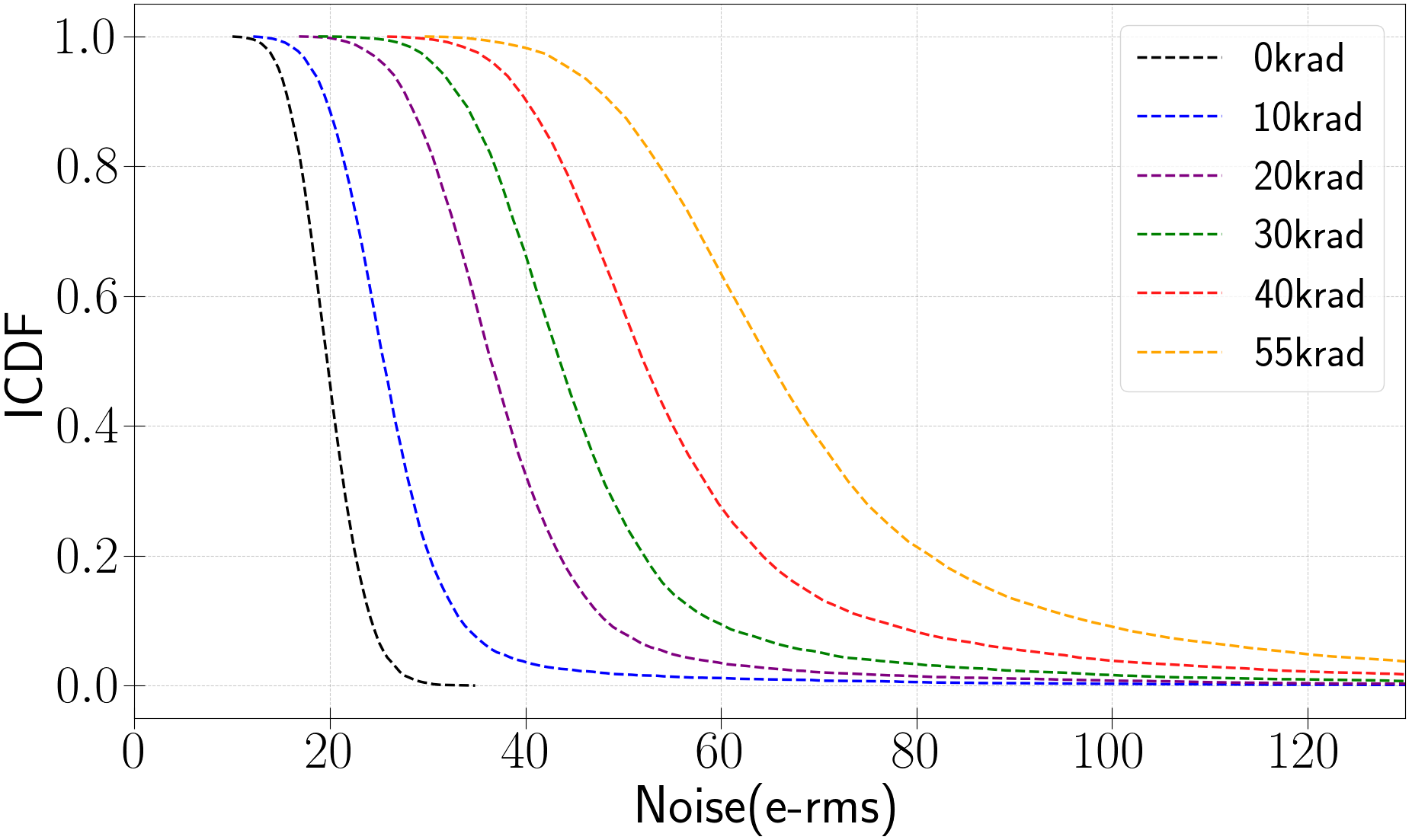}\hfill
	\caption{Inverse Cumulative Distribution Function (ICDF) of total in-pixel temporal noise for multiple TID values.}
	\label{fig:noise}
\end{figure}

Figure \ref{fig:noise} shows the distribution of temporal noise for multiple TID values. An increase in both the gaussian and especially in the non-gaussian tails of the distributions can be observed. This is an indication that the sensor's noise could be elevated due to Random Telegraph Signal (RTS). This effect was not seen after turning the TG off (TG high voltage set at $0$V), thus indicating that the RTS originates in the photo-diode.

\section{TID induced dark current Random Telegraph Signal}
Random Telegraph Signal appears as one or more meta stable noise levels with varying life times between microseconds and weeks. The leading theory is that this behavior originates at the silicon-oxide interface due to SRH combination-recombination centers \cite{5752812}. If the sensor depleted region is in contact with this interface, the charge is collected and transferred further to the floating diffusion. This means that the RTS amplitudes are proportional to the photo-diode integration time. RTS is expected to have an impact on the operation of sensors with high integration times, such as the ULTRASAT camera, which is designed with a nominal time integration of $300$ s.

\begin{figure}[h]
	\centering
	\includegraphics[width=0.81\columnwidth]
	{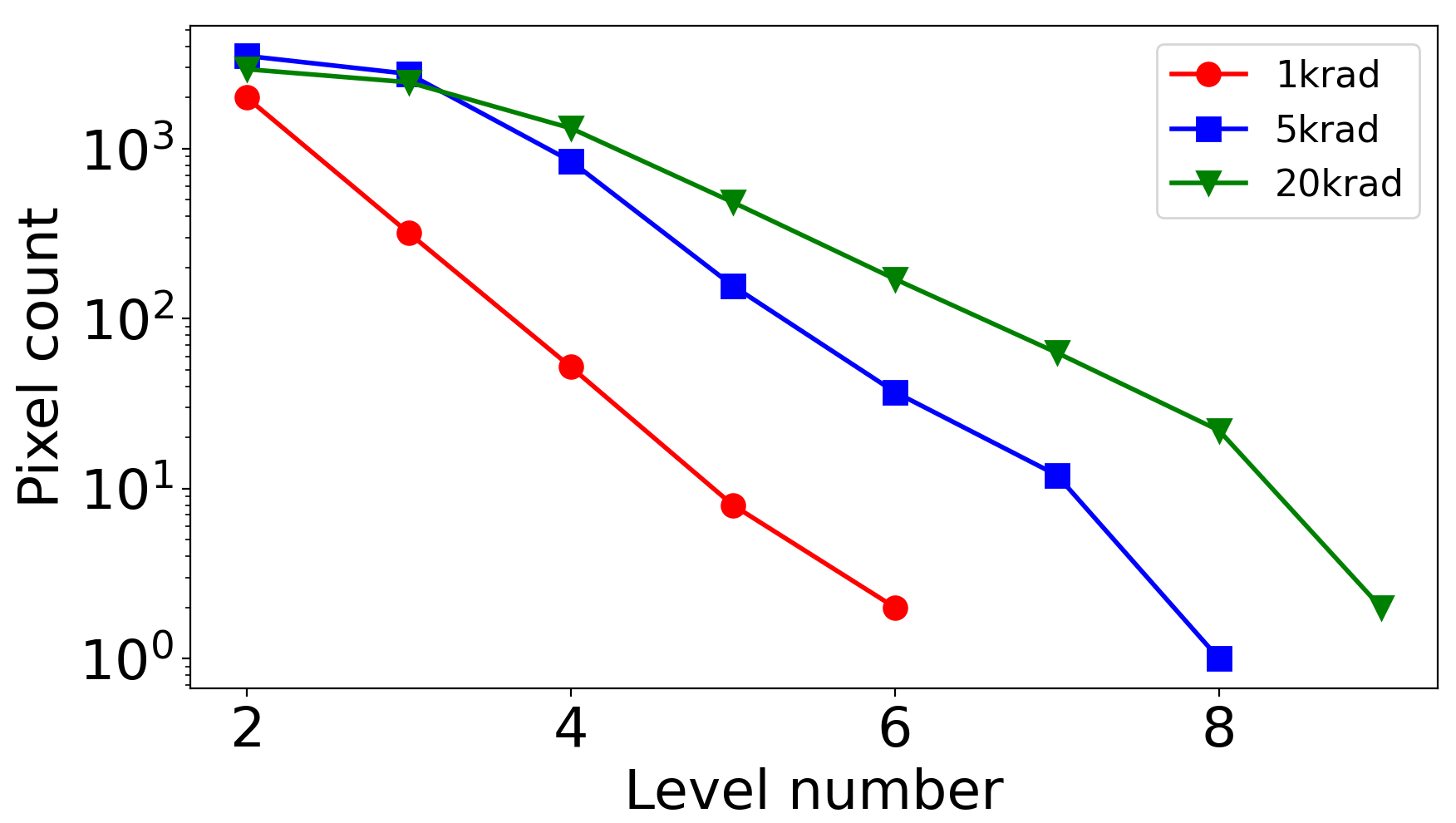}\hfill
	\caption{Distribution of the number of RTS levels for multiple TID values. The number of RTS levels is defined as the total number of noise levels identified in a pixel. Only irradiation levels with significant statistics (data) are plotted.}
	\label{fig:number}
\end{figure}

In order to characterize the magnitude of RTS in our test sensors, an automated detection technique is employed \cite{5204683}, that can identify the leading edges of the pixel signal and extract the amplitude and life time of each RTS transition. In Figure \ref{fig:number} the distribution of the number of RTS levels is plotted. For higher TID values, more RTS levels are observed \cite{6072297}. The relatively high number of pixels with multi-level RTS might be caused by the grounding of the Low TG line in our sensor design. This leads the TG to work in depletion mode, thus opening a channel between the photo-diode and the interface defects.

\begin{figure}[h]
	\centering
	\includegraphics[width=0.81\columnwidth]
	{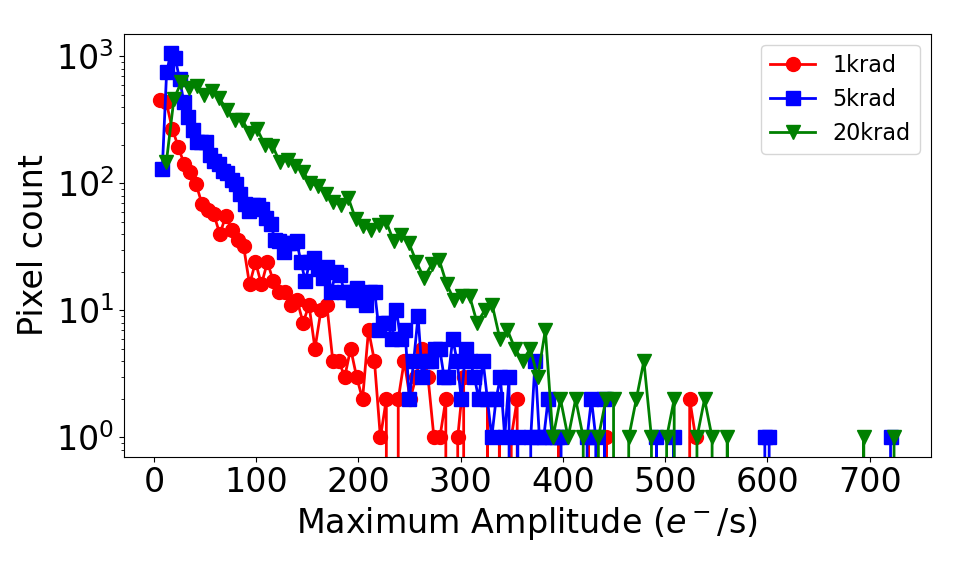}\hfill
	\caption{Distribution of the RTS maximum amplitudes for multiple RTS values. Only irradiation levels with significant statistics (data) are plotted.}
	\label{fig:ampl}
\end{figure}

One of the most important RTS parameters to analyze is the maximum amplitude distribution of RTS transitions. Figure \ref{fig:ampl} shows the maximum amplitude distribution for multiple TID values. All the curves have similar exponential decay slopes in semi-logarithmic scale, $\sim110$ $\mathrm{e^-}/s$. This hints towards the fact that the RTS mechanism is the same for all irradiation doses and is associated to TID induced damage as evidenced in \cite{5752812}.

\section{Conclusion and outlook}
ULTRASAT is planned to take low noise, high integration time images for at least 3 years in the geostationary orbit. This preliminary study indicates that RTS is a noise component that can appear during the lifetime of the camera and should be characterized. In this context, it should be mentioned that during the mission lifetime, several decontamination procedures are planned (in order to maintain a clean sensor surface). During these, the sensor's temperature will be elevated which will induce thermal annealing, that in turn is expected to revert the TID effects. 

Further tests are planned on the test structures in order to determine both the behavior of RTS in low temperature conditions close to what we expect in orbit ($200$ $K$), and also the effect of accelerated thermal annealing on the RTS. TID tests are also planned on the ULTRASAT sensor in order to verify these findings and study the impact of RTS on the sensor parameters.


\bibliography{VBerlea_arxiv_revised}

\end{document}